\documentclass[11pt, hyphens]{kimreview}
\usepackage[utf8]{inputenc} 
\usepackage[T1]{fontenc}    
\usepackage{hyperref}       
\usepackage{url}            
\usepackage{booktabs}       
\usepackage{amsfonts}       
\usepackage{nicefrac}       
\usepackage{graphicx}
\usepackage[numbers,sort&compress]{natbib}
\bibpunct[, ]{[}{]}{,}{n}{,}{,}

\usepackage{algorithm}
\usepackage{algpseudocode}

\usepackage{doi}
\usepackage{amsmath}
\usepackage{float}
\usepackage{gensymb}
\usepackage{xcolor}
\usepackage{amsthm}
\usepackage{dcolumn}
\usepackage{epsfig}
\usepackage{bm}
\usepackage{datetime}
\usepackage[title]{appendix}

\usepackage[all]{hypcap}
\begin{document}

\title{Path Sampling for Rare Events Boosted by Machine Learning}

\authorOne{Porhouy Minh}
\affiliationOne{Department of Chemistry and Chemical Theory Center, University of Minnesota Minneapolis, MN USA}

\authorTwo{Sapna Sarupria}
\affiliationTwo{Department of Chemistry and Chemical Theory Center, University of Minnesota Minneapolis, MN USA}

\publishyear{2026}
\volumenumber{4}
\articlenumber{01}
\submitdate{December 12, 2025}
\publishdate{February 04, 2026}
\doiindex{10.25950/7f47b6e6}
\doilink{10.25950/7f47b6e6}

\paperReviewed
{Machine-guided path sampling to discover mechanisms of molecular self-organization}
{Hendrik Jung, Roberto Covino, A. Arjun, Christian Leitold, Christoph Dellago, Peter G. Bolhuis and Gerhard Hummer}
{\href{https://doi.org/10.1038/s43588-023-00428-z}{Nat. Comput. Sci. 3, 334–345 (2023)}}

\maketitle

\begin{abstract}
    The study by Jung et al.~\cite{Hummer2023NCS} introduced Artificial Intelligence for Molecular Mechanism Discovery (AIMMD), a novel sampling algorithm that integrates machine learning to enhance the efficiency of transition path sampling (TPS). By enabling on-the-fly estimation of the committor probability and simultaneously deriving a human-interpretable reaction coordinate, AIMMD offers a robust framework for elucidating the mechanistic pathways of complex molecular processes. This commentary provides a discussion and critical analysis of the core AIMMD framework, explores its recent extensions, and offers an assessment of the method's potential impact and limitations.
\end{abstract}

\section*{Background}

Elucidating complex reaction mechanisms remains a central challenge in both chemical and biological sciences. Molecular dynamics (MD) simulations have long served as a powerful tool for uncovering mechanistic details, as they offer atomistic resolution and direct access to dynamical information. However, many reactions of interest are rare events that occur on timescales far longer than those accessible to straightforward MD, making direct sampling of reactive transitions prohibitively inefficient. To overcome this limitation, numerous enhanced sampling techniques have been developed, including biased approaches (e.g., umbrella sampling~\cite{Valleau1977JCP}, metadynamics~\cite{Parrinello2002PNAS,Parrinello2008PRL}) and unbiased approaches (e.g., path sampling~\cite{Geissler2002ARPC,Bolhuis2005JCP,vanErp:17:JCP,tenWolde:05:PRL,SarupriaJCP2019,RogalJCP2022}, weighted ensemble~\cite{Lillian2015JCTC}). Critically, the efficiency of these methods depends strongly on their adherence to the reaction coordinate (RC), which is unknown {\it a priori}. The RC is a special coordinate of the system that accurately captures the dynamical progress from the reactant (state A) to the product (state B) state.~\cite{Peters:16:AnnRev} The committor probability ($p_B(\mathbf{x})$), defined as the probability that a trajectory initiated from configuration $\mathbf{x}$ reaches state $B$ before returning to state $A$, is widely regarded as the optimal RC~\cite{Rogal2020JCP,Peters:16:AnnRev}. Yet, computing $p_B(\mathbf{x})$ directly is computationally demanding, as it requires shooting hundreds of short trajectories from numerous configurations. Furthermore, the $p_B$ does not directly provide any physical insights into the structural and dynamical properties important for the transition.

Transition path sampling (TPS) provides an unbiased framework for generating reactive trajectories without requiring prior knowledge of the RC. In short, TPS performs Monte Carlo (MC) moves in path space, iteratively modifying a configuration of an existing path to generate a new path. In a typical move, a configuration (also known as the shooting point) along the current reactive trajectory is selected, its momenta are perturbed, and the forward and backward integrations generate a proposed new path~\cite{Geissler2002ARPC,Dellago2010RevCompChem}. If the new trajectory connects state A to state B, it is accepted; otherwise, it is rejected. The algorithm can be summarized in Algorithm \ref{alg:cap}:
\begin{algorithm}[H]
\label{alg:tps}
\caption{Logic of the TPS Algorithm}\label{alg:cap}
\begin{algorithmic}[1]
\Require One reactive trajectory $\mathcal{TP}$ that connects state $A$ to state $B$ 

\While{sampling is not converged}

    \State Select a shooting point $\mathbf{x_s}$ from the current $\mathcal{TP}$ according to a probability $P_\text{sel}(\mathbf{x_s|\mathcal{TP}})$
    
    \State Perturb the momenta at $\mathbf{x_s}$ to $\mathbf{x_s^n}$ according to $P_\text{acc}(\mathbf{x_s} \rightarrow \mathbf{x_s^n})$
    
    \If{not \texttt{reject}}
        \While{not in B}
            \State Integrate equations of motion \textbf{forward} in time from $\mathbf{x_s^n}$ 
        
            \If{in A before in B}
                \State \texttt{reject} = True
                \State Retain the previous trajectory $\mathcal{TP}$
                \State Go back to Step 2
            \EndIf
        \EndWhile

        \While{not in A}
            \State Integrate equations of motion \textbf{backward} in time from $\mathbf{x_s^n}$
            
            \If{in B before in A}
                \State \texttt{reject} = True
                \State Retain the previous trajectory $\mathcal{TP}$
                \State Go back to Step 2
            \EndIf
        \EndWhile
    
        \State Combine the forward and backward segments and record the new trajectory $\mathcal{TP}'$
    
        \State  Update $\mathcal{TP} \gets \mathcal{TP}'$ 
   
    \EndIf
    
\EndWhile
\end{algorithmic}
\end{algorithm}

Although TPS bypasses the need for the RC when sampling, its efficiency still depends on choosing a reasonable collective variable (CV)---a physical quantity of the system---that can distinguish between states A and B. TPS can suffer from poor acceptance rates or inadequate exploration when the chosen CV does not correlate well with the progress of the reaction. Thus, identifying an effective CV or the RC remains a major bottleneck.

Artificial Intelligence for Molecular Mechanism Discovery (AIMMD) addresses this bottleneck by integrating machine learning (ML) with the TPS framework. Rather than relying on a fixed CV, AIMMD learns the RC as the committor on-the-fly while performing the ML-enhanced TPS. This on-the-fly learning and adaptation of the RC effectively mitigates the issues associated with a poorly chosen sampling coordinate, allowing the algorithm to focus computational resources on the transition region, which is critical for generating transition paths (TP) in TPS. Furthermore, AIMMD elevates the scientific utility of the committor by employing symbolic regression to describe it as an analytical function of preselected CVs, thereby expanding its interpretability. Overall, AIMMD showcases a robust framework for the efficient discovery and characterization of complex molecular pathways. 

\section*{Overview of AIMMD}
The AIMMD algorithm is initiated in a manner analogous to TPS. First, states A and B of the system are defined along a physically motivated CV. An initial TP connecting A and B is then generated to jump-start TPS. After performing several TPS moves, the shooting points and their outcomes from this initial run provide the training data for the first ML cycle.

\vspace{\baselineskip}
\begin{figure}[h!]
    \centering
    \includegraphics[width=\textwidth]{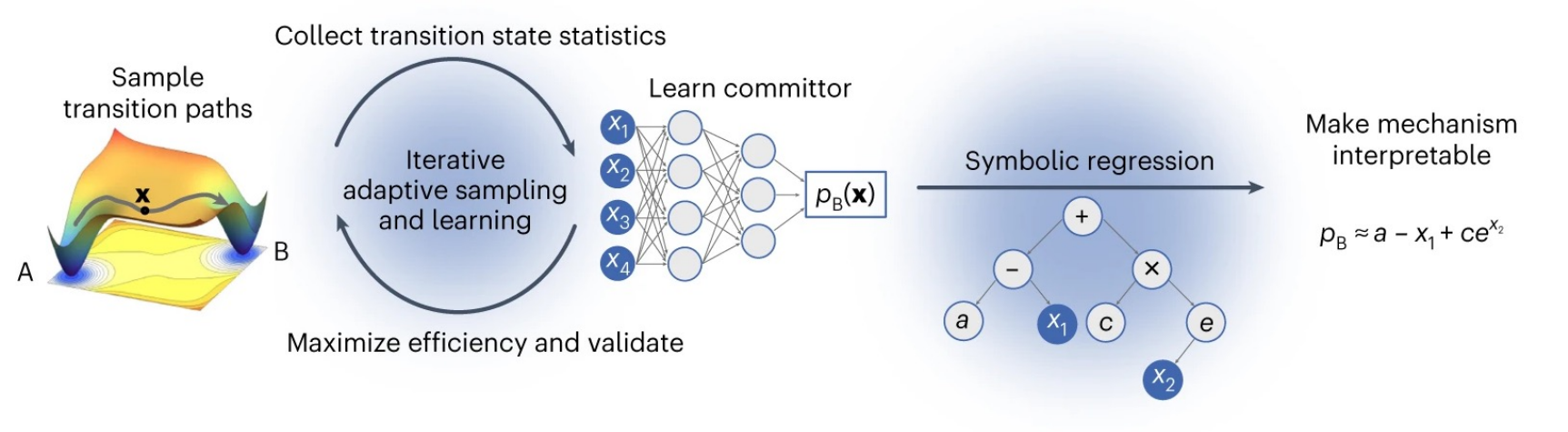}
    \caption{\textbf{Overview of AIMMD Schematic.} First, AIMMD is trained in a self-consistent manner, where the process alternates between rounds of sampling using TPS and training the neural network. Once trained, the resulting committor function is described in terms of the physical CVs of the system through symbolic regression. Figure reproduced with permission from Figure 1a of Ref.~\citenum{Hummer2023NCS}.}
    \label{fig:aimmd-scheme}
\end{figure}
\vspace{\baselineskip}

AIMMD employs a feed-forward neural network in which each data point $\mathbf{x}$ corresponds to a molecular configuration (i.e., a shooting point) described by $N$ physical CVs (i.e., features). For example, in an ion association--dissociation process, such CVs may include the inter-ionic distance or the coordination numbers of surrounding solvent atoms. These inputs are nonlinearly combined as they propagate through the network to output the RC in the form of a logit-committor, $q(\mathbf{x}|\boldsymbol{\theta})$. The committor is then given by Eq.~\ref{eq:pb}:
\begin{equation}
p_B(\mathbf{x})=\frac{1}{1+e^{-q(\mathbf{x}|\boldsymbol{\theta})}}
\label{eq:pb}
\end{equation}

\noindent The network parameters $\boldsymbol{\theta}$ are learned from the successes and failures of the most recent $k$ shooting attempts. Training is performed by minimizing the negative log-likelihood loss, 
\begin{equation}
L=-\sum_{i=1}^{k}\left[n_i^A\ln (1-p_B(\mathbf{x}_i)) + n_i^B\ln p_B(\mathbf{x}_i)\right] = -\sum_{i=1}^{k}\left[n_i^A\ln (1-\frac{1}{1+e^{-q(\mathbf{x}|\boldsymbol{\theta})}}) - n_i^B\ln(1+e^{-q(\mathbf{x}|\boldsymbol{\theta})})\right]
\label{eq:logloss}
\end{equation}
where $n_i^A$ and $n_i^B$ encode the outcome of shooting point $\mathbf{x}_i$. For a reactive path (AB), $n_i^A=n_i^B=1$. For nonreactive paths, $n_i^A=2, n_i^B=0$ for AA trajectories and $n_i^A=0, n_i^B=2$ for BB trajectories. Minimizing this loss, therefore, yields network parameters that maximize the likelihood of the observed shooting outcomes.

AIMMD proceeds in a self-consistent loop, where TPS sampling and ML training alternate (Fig. \ref{fig:aimmd-scheme}). After each round of model training, TPS continues by selecting a shooting point on the most recently generated TP according to a Lorentzian distribution centered on the transition-state ensemble (TSE):
\begin{equation}
P_{\text{sel}}(\mathbf{x}|\mathcal{TP})=
\frac{1}{\sum_{\mathbf{x}'\in\mathcal{TP}}
\left[\frac{q(\mathbf{x})^2+\gamma^2}{q(\mathbf{x}')^2+\gamma^2}\right]},
\label{eq:lorentzian}
\end{equation}
where larger values of $\gamma$ promote broader exploration. This scheme enhances sampling efficiency by favoring shooting points near the transition state, while still permitting occasional excursions away from it.

The training cycle terminates once the committor model is deemed sufficient, as determined by whether the expected number of TPs from the last $k$ shooting points matches the number actually observed. The expected number of TPs is given by
\begin{equation}
n_{\mathcal{TP}}^{\text{exp}}=\sum_{i=1}^{k} 2\bigl(1-p_B(\mathbf{x}_i)\bigr) p_B(\mathbf{x}_i)
\label{eq:n_TP_exp}
\end{equation}
where $p_B(\mathbf{x}_i)$ is the predicted probability of reaching state B before the shooting outcome at step $i$ is known. Once trained, the committor can be expressed in an interpretable form using symbolic regression, which represents the learned RC as a combination of known physical CVs. Although the degree of interpretability depends on how complex the resulting functional form is, this representation still provides insights into the relative importance and interactions of the underlying CVs of the system.

In Jung \textit{et al.}\cite{Hummer2023NCS}, AIMMD was applied to a range of systems, including ion association/dissociation in solution, gas hydrate nucleation, homopolymer folding, and the assembly of the transmembrane protein Mga2 in a lipid bilayer. Notably, in the case of ion association/dissociation, Jung \textit{et al.}~\cite{Hummer2023NCS} demonstrated transferability of the learned mechanism. Starting from a model trained for LiCl, they performed transfer learning by retraining only the final neural network layer to study the ion association-dissociation of other monovalent salts. In the Mga2 assembly study, they further showcased the use of parallel TPS simulations to generate pooled data for model training. This strategy enabled the identification of two distinct reaction pathways and produced a committor landscape that reflects their coexistence.

\section*{Commentary and Critical Analysis}

While AIMMD presents a promising approach for automated RC discovery and efficient rare event sampling, our examination highlights several areas where methodological rigor and further development would strengthen the framework. The method clearly demonstrates its ability to identify an RC that correlates with the committor, thereby facilitating the localization of the TSE and improving the efficiency of generating TPs within TPS. However, a notable gap remains regarding a rigorous assessment of the quality of the learned committor. Although Jung \textit{et al.} showed that the predicted committor aligns well with numerically estimated committor values, the framework itself does not incorporate a definitive procedure for verifying that the learned RC is the true committor. A well-established validation tool---the histogram test~\cite{Peters2006JCP,Peters:16:AnnRev}---provides such an assessment by checking whether configurations predicted to have the same committor indeed exhibit statistically identical outcomes in short shooting trials. Integrating this test into AIMMD would provide a quantitative evaluation of whether the learned function accurately represents the true committor.

A second issue concerns the sampling imbalance inherent in the learning procedure. Although AIMMD employs a Lorentzian distribution for $P_{\mathrm{sel}}$ to balance exploration and exploitation, the majority of shooting points chosen would be configurations near the transition state, which generates TPs. Consequently, the training data are disproportionately concentrated around configurations with $p_B(\mathbf{x}) \approx 0.5$. This uneven sampling reduces the predictive accuracy of the learned committor for regions farther from the TSE (i.e., for configurations with $p_B < 0.5$ or $p_B > 0.5$), where fewer data points are available. In principle, this could hinder downstream tasks that rely on accurate committor predictions away from the TSE. Therefore, highlighting strategies to adaptively broaden the sampling distribution or to include additional exploratory shooting moves would enhance the robustness of the approach.

Furthermore, AIMMD inherits a key limitation of TPS itself, which is the difficulty in thoroughly exploring multiple reaction pathways when the system exhibits branching mechanisms. Accessing distinct pathways requires generating a shooting point near the dividing region separating the pathways, which may be rare if the energy barrier between pathways is non-negligible. As a result, the probability of discovering new pathways depends sensitively on both the barrier height and the degree of overlap between the respective TSEs. In contrast, methods such as replica exchange transition interface sampling (RETIS)~\cite{vanErp:17:JCP} employ exchanges between neighboring interface ensembles, allowing trajectories to cross barriers in "pathway space" more readily. AIMMD partially addresses this limitation through training on data from multiple, parallel TPS instances, as demonstrated in the Mga2 study. However, the degree to which this strategy mitigates sampling bias depends strongly on how those parallel trajectories are initialized. A more detailed discussion of how AIMMD performs when confronted with reaction networks containing multiple, well-separated pathways would benefit future users aiming to apply the method in such settings.

\section*{Extensions of AIMMD}

The AIMMD framework has already motivated several methodological extensions that broaden its utility and scope. An early extension of AIMMD is the waste-recycling TPS scheme introduced by Lazzeri \textit{et al.}~\cite{Bolhuis2023JCTC}. In standard TPS, only successful TPs are used for analysis, whereas nonreactive (AA and BB) trajectories are discarded. The waste-recycling variant utilizes these rejected paths, together with data from additional equilibrium simulations, to extract information from a more extensive sampling of the configuration space. Incorporating this broader dataset yields a more complete characterization of the reaction landscape, enabling improved estimation of both the rate constant and the free energy surface as a function of the learned committor. Furthermore, in this AIMMD-TPS implementation, shooting points are chosen to balance the \textit{exploration} of the relevant configurational region with \textit{exploitation} of points likely to generate reactive trajectories by using a uniform distribution as a function of the committor instead of the Lorentzian. This strategy helps alleviate the imbalance in training data that arises when most samples originate from near the transition region, thereby improving the committor estimates for configurations farther from the transition state.

A more recent extension integrates AIMMD with Transition Interface Sampling (AIMMD-TIS)~\cite{Bolhuis2025arXiv}. In this approach, an initial AIMMD-TPS run provides a preliminary estimate of the logit-committor $q(\mathbf{x})$, which is then used to determine approximate boundaries for states A and B and to place TIS interfaces that serve as analogs of isocommittor surfaces. Since TIS saves both reactive and nonreactive trajectories, it generates a much more diverse ensemble of configurations than TPS alone. This expanded sampling substantially improves the coverage of the committor across configuration space, particularly in regions away from the TSE where AIMMD-TPS may lack sufficient data. At convergence, AIMMD-TIS yields not only the reaction rate and free energy surface, but also a refined committor model and an associated set of isocommittor surfaces. The latter enables the computation of the mean gradient of the logit-committor with respect to each neural network input feature, thereby providing a systematic and quantitative measure of feature importance at different stages along the reaction pathway.

Furthermore, by leveraging the interface structure of TIS, AIMMD-TIS enhances the exploration of alternative reaction pathways compared to AIMMD-TPS. The use of multiple interfaces, each generating their own ensemble of trajectories, increases the likelihood of sampling distinct mechanistic channels. While this makes AIMMD-TIS a better choice for systems with multiple pathways or mechanistic heterogeneity compared to AIMMD-TPS, an approach incorporating RETIS with AIMMD would provide an even more comprehensive exploration of pathway diversity.

\section*{Conclusion}

Historically, the broader adoption of path sampling methods has lagged behind that of biased techniques such as metadynamics, in part due to limited software accessibility. This gap is now narrowing with the development of user-friendly tools such as OpenPathSampling (OPS)~\cite{ops1,ops2}, AIMMD’s own software package~\cite{Hummer2023NCS}, and PyRETIS~\cite{PyRETIS1,PyRETIS2,PyRETIS3}, which collectively lower the barrier to entry for practitioners.

A further practical challenge arises from the inherently iterative nature of TPS-based approaches. For highly diffusive processes, such as salt nucleation, reactive trajectories can be extremely long, which raises the question about the efficiency of repeatedly generating such trajectories during RC refinement. The extent to which AIMMD can maintain its efficiency in these slow dynamical regimes remains to be fully tested. Nevertheless, if the RC can be refined significantly using only a modest number of trajectories, as demonstrated in the ion association–dissociation case through few-shot learning, AIMMD has the potential to remain highly effective even in complex, slow dynamical systems.

Overall, AIMMD serves as a valuable tool for computational chemistry, providing an automated and interpretable framework for sampling and elucidating complex pathways. By combining ML with TPS, AIMMD not only enhances sampling efficiency but also yields physically meaningful, interpretable RC through symbolic regression. Through directly addressing one of the most pressing computational bottlenecks in simulations, it has the potential to rapidly enhance the understanding of complex systems. 

\section*{Acknowledgments}
 This work was supported by the U.S. Department of Energy, Office of Science, Office of Basic Energy Sciences, under Award No. DE-SC0025496. P.M. acknowledges support from Wayland E. Noland via the Wayland E. Noland Excellence Fellowship in Chemistry.

\end{document}